\documentclass[conference]{IEEEtran}
\IEEEoverridecommandlockouts

\usepackage{amsmath,amsfonts}
\usepackage{algorithmic}
\usepackage{array}
\usepackage[caption=false,font=normalsize,labelfont=sf,textfont=sf]{subfig}
\usepackage{textcomp}
\usepackage{stfloats}
\usepackage{url}
\usepackage{verbatim}
\usepackage{graphicx}
\def\BibTeX{{\rm B\kern-.05em{\sc i\kern-.025em b}\kern-.08em
    T\kern-.1667em\lower.7ex\hbox{E}\kern-.125emX}}
\usepackage{balance}
\usepackage{caption}
\usepackage{siunitx}
\usepackage{tikz}
\usetikzlibrary{shapes.geometric, arrows.meta, positioning, calc, fit}

\usepackage{algorithm}
\usepackage{mathtools}	
\usepackage{amssymb}
\usepackage{amsthm}

\usepackage[utf8]{inputenc}

\usepackage{ifthen}

\usepackage{microtype}

\usepackage{caption}

\usepackage{bm}

\usepackage[draft,nomargin,marginclue,inline]{fixme}
\makeatletter
\renewcommand*\FXLayoutMarginClue[3]{%
  \marginpar[%
  \raggedleft\@fxuseface{margin}\textcolor{red}{\ignorespaces $ \Rightarrow $}]{%
    \raggedright\@fxuseface{margin}\textcolor{red}{\ignorespaces $ \Leftarrow $}}}
\makeatother	

\usepackage{tumcolor}

\usepackage{pgfplotstable}	

\pgfplotsset{
	discard if/.style 2 args={
        x filter/.append code={
            \edef\tempa{\thisrow{#1}}
            \edef\tempb{#2}
            \ifx\tempa\tempb
                
            \fi
        }
    },
    discard if not/.style 2 args={
        x filter/.append code={
            \edef\tempa{\thisrow{#1}}
            \edef\tempb{#2}
            \ifx\tempa\tempb
            \else
                
            \fi
        }
    }
}

\usepackage{cite}

\usepackage[acronym,shortcuts]{glossaries}
\newacronym{cnn}{CNN}{convolutional neural network}
\newacronym{ula}{ULA}{uniform linear array}

\usepackage{cleveref}

\tikzset{algorithm1/.style={mark options={solid},color=TUMBeamerBlue, line width=\lineWidth, mark=square, dashed}}

\pgfplotscreateplotcyclelist{miko}{%
{color=TUMBeamerRed, dash pattern=on 5pt off 1pt on 1pt off 1pt, line width=1.5pt},%
{color=TUMBeamerOrange, dash pattern=on 5pt off 2pt, line width=1.5pt,
mark=o},%
{color=black, dash pattern=on 5pt off 2pt on 3pt off 2pt, line width=1.5pt},%
{color=TUMDarkerBlue, line width=1.5pt},%
{color=black, dotted, line width=1.5pt},%
{color=TUMBeamerGreen, dotted, line width=1.5pt},%
{color=TUMBeamerYellow, dotted, line width=1.5pt},%
{color=TUMBeamerDarkRed, dotted, line width=1.5pt},%
{color=TUMBeamerLightGreen, dotted, line width=1.5pt},%
{color=TUMIvory, dotted, line width=1.5pt},%
{color=TUMLightBlue, dotted, line width=1.5pt},%
}

\DeclareMathOperator*{\argmax}{arg\,max}

\DeclareMathOperator{\diag}{diag}

\DeclareMathOperator{\expec}{E}

\newcommand{\calN}{\mathcal{N}}
\newcommand{\calO}{\mathcal{O}}

\newcommand*{\C}{\mathbb{C}}

\newcommand*{\R}{\mathbb{R}}

\newcommand{\herm}{{\operatorname{H}}}
\newcommand{\tp}{{\operatorname{T}}}

\definecolor{myblue}{RGB}{30, 100, 200}

\newlength{\leftstackrelawd}
\newlength{\leftstackrelbwd}
\def\leftstackrel#1#2{\settowidth{\leftstackrelawd}%
	{${{}^{#1}}$}\settowidth{\leftstackrelbwd}{$#2$}%
	\addtolength{\leftstackrelawd}{-\leftstackrelbwd}%
	\leavevmode\ifthenelse{\lengthtest{\leftstackrelawd>0pt}}%
	{\kern-.5\leftstackrelawd}{}\mathrel{\mathop{#2}\limits^{#1}}}

\newcommand{\mbC}{\bm{C}}
\newcommand{\mbD}{\bm{D}}

\newcommand{\mbF}{\bm{F}}

\newcommand{\mbP}{\bm{P}}

\newcommand{\mbX}{\bm{X}}

\newcommand{\mbc}{\bm{c}}

\newcommand{\mbh}{\bm{h}}

\newcommand{\mbn}{\bm{n}}

\newcommand{\mbp}{\bm{p}}

\newcommand{\mbv}{\bm{v}}

\newcommand{\mbx}{\bm{x}}
\newcommand{\mby}{\bm{y}}

\newcommand{\mbSigma}{{\bm{\Sigma}}}

\newcommand{\mbmu}{{\bm{\mu}}}

\newcommand{\hhat}{\hat{\mbh}}

\newcommand{\covhi}{\mbC_i}
\newcommand{\covhk}{\mbC_k}
\newcommand{\meanhi}{\mbmu_i}
\newcommand{\meanhk}{\mbmu_k}

\usetikzlibrary{intersections,pgfplots.fillbetween,patterns,spy}

\Crefname{figure}{Fig.}{Figs.}
\pgfplotsset{compat=1.15}

\ifdefined\ONECOLUMN

\newcommand{\normalplotwidth}{0.5\columnwidth}
\newcommand{\largeplotheight}{0.36\columnwidth}
\else

\newcommand{\normalplotwidth}{1.0\columnwidth}
\newcommand{\largeplotheight}{0.65\columnwidth}
\newcommand{\largeplotwidth}{1.0\columnwidth}
\fi

\newcommand{\lineWidth}{1.0pt}
\newcommand{\markSize}{2.0pt}
\definecolor{ourdarkblue}{RGB}{30, 100, 200}
\definecolor{ourdarkgreen}{RGB}{0, 100, 0}
\definecolor{ourdarkorange}{RGB}{201, 98, 18}
\definecolor{ouryellow}{RGB}{220, 210, 50}

\tikzset{unipowcdf/.style={mark options={solid}, color=ourdarkgreen, line width=\lineWidth, mark=None, mark size=\markSize, dotted}}
\tikzset{upchansubcdf/.style={mark options={solid}, color=purple, line width=\lineWidth, mark=None, mark size=\markSize, dotted}}

\tikzset{lloydpgddl/.style={mark options={solid}, color=TUMBeamerOrange, line width=\lineWidth, mark=None, mark size=\markSize}}
\tikzset{lloydpgdul/.style={mark options={solid}, color=cyan, line width=\lineWidth, mark=None, mark size=\markSize, dotted}}
\tikzset{lloydlaudl/.style={mark options={solid}, color=TUMBeamerRed, line width=\lineWidth, mark=None, mark size=\markSize}}
\tikzset{lloydlauul/.style={mark options={solid}, color=TUMBeamerGreen, line width=\lineWidth, mark=None, mark size=\markSize, dotted}}
\tikzset{gmmpgddl/.style={mark options={solid}, color=TUMBlue, line width=\lineWidth, mark=None, mark size=\markSize}}
\tikzset{gmmpgdul/.style={mark options={solid}, color=brown, line width=\lineWidth, mark=None, mark size=\markSize, dotted}}
\tikzset{gmmlaudl/.style={mark options={solid}, color=gray, line width=\lineWidth, mark=None, mark size=\markSize}}
\tikzset{gmmlauul/.style={mark options={solid}, color=black, line width=\lineWidth, mark=None, mark size=\markSize, dotted}}

\tikzset{gmmmrgmrg/.style={mark options={solid}, color=gray, line width=\lineWidth, mark=None, mark size=\markSize}}

\tikzset{lloydpgdulperfect/.style={mark options={solid}, color=gray, line width=\lineWidth, mark=None, mark size=\markSize, dashed}}
\tikzset{lloydpgdulomp/.style={mark options={solid}, color=TUMBeamerOrange, line width=\lineWidth, mark=None, mark size=\markSize, dashed}}
\tikzset{lloydpgdulls/.style={mark options={solid}, color=red, line width=\lineWidth, mark=None, mark size=\markSize}}
\tikzset{lloydpgdulscov/.style={mark options={solid}, color=TUMBeamerRed, line width=\lineWidth, mark=None, mark size=\markSize, dashed}}
\tikzset{lloydpgdulgmmulbest/.style={mark options={solid}, color=gray, line width=\lineWidth, mark=None, mark size=\markSize}}
\tikzset{lloydpgdulgmmulall/.style={mark options={solid}, color=black, line width=\lineWidth, mark=None, mark size=\markSize, dashed}}

\tikzset{gmmpgdulperfect/.style={mark options={solid}, color=TUMBeamerBlue, line width=\lineWidth, mark=None, mark size=\markSize}}
\tikzset{gmmpgdulfromy/.style={mark options={solid}, color=TUMBeamerGreen, line width=\lineWidth, mark=None, mark size=\markSize}}

\tikzset{MUlloydpgdulperfect/.style={mark options={solid}, color=violet, line width=\lineWidth, mark=None, mark size=\markSize, dashed}}
\tikzset{MUlloydpgdulomp/.style={mark options={solid}, color=TUMBeamerOrange, line width=\lineWidth, mark=None, mark size=\markSize, dashed}}
\tikzset{MUlloydpgdulscov/.style={mark options={solid}, color=TUMBeamerRed, line width=\lineWidth, mark=None, mark size=\markSize, dashed}}
\tikzset{MUlloydpgdulgmmulall/.style={mark options={solid}, color=black, line width=\lineWidth, mark=None, mark size=\markSize, dashed}}

\tikzset{MUrandomperfect/.style={mark options={solid}, color=black, line width=\lineWidth, mark=None, mark size=\markSize, dotted}}
\tikzset{MUrandomomp/.style={mark options={solid}, color=TUMBeamerOrange, line width=\lineWidth, mark=None, mark size=\markSize, dotted}}
\tikzset{MUrandomscov/.style={mark options={solid}, color=TUMBeamerRed, line width=\lineWidth, mark=None, mark size=\markSize, dotted}}
\tikzset{MUrandomgmmulall/.style={mark options={solid}, color=gray, line width=\lineWidth, mark=None, mark size=\markSize, dotted}}

\tikzset{MUgmmpgdulperfect/.style={mark options={solid}, color=TUMBeamerBlue, line width=\lineWidth, mark=None, mark size=\markSize}}
\tikzset{MUgmmpgdulfromy/.style={mark options={solid}, color=TUMBeamerGreen, line width=\lineWidth, mark=None, mark size=\markSize}}

\tikzset{MUswmmsegmmpgdulperfect/.style={mark options={solid}, color=TUMBeamerBlue, line width=\lineWidth, mark=None, mark size=\markSize, dashdotted}}
\tikzset{MUswmmsegmmpgdulfromy/.style={mark options={solid}, color=TUMBeamerGreen, line width=\lineWidth, mark=None, mark size=\markSize, dashdotted}}

\newcommand{\pltMUdftperfect}{DFT, $\mbh$}
\newcommand{\pltMUdftomp}{DFT, $\hhat_{\text{OMP}}$}
\newcommand{\pltMUdftscov}{DFT, $\hhat_{\text{LMMSE}}$}
\newcommand{\pltMUdftgmmulall}{DFT, $\hhat_{\text{GMM}}$}

\newacronym{AWGN}{AWGN}{additive white Gaussian noise}
\newacronym{BLMMSE}{BLMMSE}{Bussgang LMMSE}
\newacronym{BS}{BS}{base station}
\newacronym{CDF}{CDF}{cumulative distribution function}
\newacronym{CNN}{CNN}{convolutional neural network}
\newacronym{CSI}{CSI}{channel state information}
\newacronym{CSIT}{CSIT}{channel state information at the transmitter}
\newacronym{DFT}{DFT}{Discrete Fourier transform}
\newacronym{DL}{DL}{downlink}
\newacronym{DNN}{DNN}{deep neural network}
\newacronym{DoA}{DoA}{direction of arrival}
\newacronym{EM}{EM}{expectation maximization}
\newacronym{FDD}{FDD}{frequency division duplex}
\newacronym{GMM}{GMM}{Gaussian mixture model}
\newacronym{LMMSE}{LMMSE}{linear minimum mean square error}
\newacronym{LOS}{LOS}{line of sight}
\newacronym{LS}{LS}{least squares}
\newacronym{MSE}{MSE}{mean squared error}
\newacronym{MIMO}{MIMO}{multiple-input multiple-output}
\newacronym{MPC}{MPC}{multi-path component}
\newacronym{MT}{MT}{mobile terminal}
\newacronym{NLOS}{NLOS}{non-line of sight}
\newacronym{NN}{NN}{neural network}
\newacronym{O2I}{O2I}{outdoor-to-indoor}
\newacronym{OMP}{OMP}{orthogonal matching pursuit}
\newacronym{PDF}{PDF}{probability density function}
\newacronym{PGA}{PGA}{projected gradient ascent}
\newacronym{PSD}{PSD}{power spectral density}
\newacronym{SNR}{SNR}{signal-to-noise ratio}
\newacronym{TDD}{TDD}{time division duplex}
\newacronym{UL}{UL}{uplink}
\newacronym{ULA}{ULA}{uniform linear array}
\newacronym{URA}{URA}{uniform rectangular array}
\newacronym{UMa}{UMa}{urban macrocell}
\newacronym{nSE}{nSE}{normalized spectral efficiency}
\newacronym{cCDF}{cCDF}{complementary cumulative distribution function}
\newacronym{MU-MIMO}{MU-MIMO}{multi-user MIMO}
\newacronym{MU-MISO}{MU-MISO}{multi-user MISO}
\newacronym{BD}{BD}{block diagonalization}
\newacronym{RBD}{RBD}{regularized block diagonalization}
\newacronym{RCI}{RCI}{regularized channel inversion}
\newacronym{WMMSE}{WMMSE}{weighted minimum mean square error}
\newacronym{SWMMSE}{SWMMSE}{stochastic WMMSE}
\newacronym{SVD}{SVD}{singular value decomposition}
\newacronym{SR}{SR}{sum-rate}
\newacronym{CME}{CME}{conditional mean estimator}
\newacronym{ML}{ML}{machine learning}
\newacronym{FLOPS}{FLOPS}{floating-point operations}
\newacronym{OFDM}{OFDM}{orthogonal frequency-division multiplexing}
\newacronym{LTE}{LTE}{Long Term Evolution}
\newacronym{GPS}{GPS}{Global Positioning System}
\newacronym{UMi}{UMi}{urban microcell}

\newcommand{\Ntx}{N_{\mathrm{tx}}}
\newcommand{\Nv}{N_{\mathrm{v}}}
\newcommand{\Nh}{N_{\mathrm{h}}}

\begin{document}



\title{Limited Feedback on Measurements:\\ Sharing a Codebook or a Generative Model?}

\author{Nurettin~Turan\IEEEauthorrefmark{1}, Benedikt~Fesl\IEEEauthorrefmark{1}, Michael Joham\IEEEauthorrefmark{1}, Zhengxiang Ma\IEEEauthorrefmark{2}, Anthony C. K. Soong\IEEEauthorrefmark{2},\\ 
Baoling Sheen\IEEEauthorrefmark{2},
Weimin Xiao\IEEEauthorrefmark{2},
and Wolfgang~Utschick\IEEEauthorrefmark{1}\\
\IEEEauthorblockA{\IEEEauthorrefmark{1}TUM School of Computation, Information and Technology, Technical University of Munich, Germany\\
 \IEEEauthorrefmark{2}Futurewei Technologies, Bridgewater, New Jersey, USA\\
	Email: nurettin.turan@tum.de
    }

\thanks{\IEEEauthorrefmark{1}The authors acknowledge the financial support by the Federal Ministry of
Education and Research of Germany in the program of ``Souver\"an. Digital.
Vernetzt.''. Joint project 6G-life, project identification number: 16KISK002. 
}
\thanks{\copyright This work has been submitted to the IEEE for possible publication. Copyright may be transferred without notice, after which this version may no longer be accessible.
}
}


\maketitle

\begin{abstract}
\ac{DFT} codebook-based solutions are well-established for limited feedback schemes in \ac{FDD} systems. 
In recent years, data-aided solutions have been shown to achieve higher performance, enabled by the adaptivity of the feedback scheme to the propagation environment of the \ac{BS} cell.
In particular, a versatile limited feedback scheme utilizing \acp{GMM} was recently introduced.
The scheme supports multi-user communications, exhibits low complexity, supports parallelization, and offers significant flexibility concerning various system parameters.
Conceptually, a \ac{GMM} captures environment knowledge and is subsequently transferred to the \acp{MT} for online inference of feedback information.
Afterward, the \ac{BS} designs precoders using either directional information or a generative modeling-based approach.
A major shortcoming of recent works is that the assessed system performance is only evaluated through synthetic simulation data that is generally unable to fully characterize the features of real-world environments. 
It raises the question of how the {GMM}-based feedback scheme performs on real-world measurement data, especially compared to the well-established \ac{DFT}-based solution. 
Our experiments reveal that the \ac{GMM}-based feedback scheme tremendously improves the system performance measured in terms of sum-rate, allowing to deploy systems with fewer pilots or feedback bits.

\end{abstract}


\begin{IEEEkeywords}
Gaussian mixture models, machine learning, limited feedback, precoding, measurement data.
\end{IEEEkeywords}

\section{Introduction}

In the upcoming generation of cellular communication systems (6G), the \ac{BS} can adapt to dynamic channel conditions.
In \ac{FDD} systems, this adaptation relies on \ac{CSI} feedback from the \acp{MT} due to the absence of channel reciprocity~\cite{Love}.
In particular, limited feedback systems characterized by only a few bits representing the feedback information are of significant interest~\cite{Love}.
In this regard, two primary approaches can be identified.
The first entails estimating the \ac{DL} channel at the \acp{MT} and subsequently determining the feedback information using a codebook~\cite{Love, KaKoGeKn09}. 
In scenarios with spatial correlation due to specific antenna geometries, \ac{DFT}-based codebooks are well-established and are part of 3GPP specifications \cite{YaYaHa10, LiSuZeZhYuXiXu13, 3GPP_5G}.
The second approach aims to circumvent channel estimation and instead encode feedback information directly from pilot observations, enabled through data-aided approaches, e.g., deep learning~\cite{GuWeChJi22, JaLeKiLe22, TuKoBaXuUt21}.

A promising approach from the latter category, albeit it does not employ deep learning, is a versatile limited feedback scheme based on \acp{GMM}~\cite{TuFeKoJoUt23}.
This scheme offers flexibility with respect to various system parameters such as the number of served \acp{MT}, the number of pilots, various \ac{SNR} levels, and the selection of the precoding algorithm.
The scheme was further enhanced to support variable feedback bit lengths in \cite{TuFeUt23}.
\acp{GMM} are generative models with a discrete latent space.
Conceptually, a \ac{GMM} is utilized to learn the underlying channel distribution of the environment of a \ac{BS} cell. 
The learned \ac{GMM} is then transferred to all \acp{MT} within the coverage area of the \ac{BS}, enabling the \acp{MT} to exploit the environment knowledge during the online phase to infer feedback information.
The computational complexity associated with the inference of the feedback index does not depend on the number of transmit antennas, which is particularly beneficial for massive \ac{MIMO} systems.
Following the \ac{MT}'s feedback report to the \ac{BS}, the \ac{BS} designs precoders using either a directional information-based method or a generative modeling-based approach.

Although the \ac{GMM}-based limited feedback scheme is a promising candidate for supporting future generations of cellular systems (6G), it is yet unclear how the performance gain that was observed for synthetic data, e.g., through stochastic-geometric channel models, transfers to real-world systems. 
Thereby, a comparison to the well-established \ac{DFT} codebook-based feedback scheme is of particular practical importance. 
In this work, we address the question of whether sharing the knowledge of a generative model, i.e., the \ac{GMM}, at the \ac{BS} and the \acp{MT} is advantageous as compared to shared \ac{DFT}-based codebook knowledge by evaluating and comparing the two feedback schemes on real-world data from a measurement campaign.
Our analyses reveal that the \ac{GMM}-based feedback scheme, which is tailored to the \ac{BS} cell environment, provides a robust solution with huge performance gains as compared to the \ac{DFT}-based codebook solution, especially in systems with low pilot overhead that are of crucial importance in massive \ac{MIMO} systems\cite{BjLaMa16}.
The performance gains offered by the \ac{GMM}-based scheme can be exploited to deploy systems with fewer pilots and feedback bits.

\section{System Model and Channel Data}
\label{sec:system_channel_model}

\subsection{Data Transmission Phase}

We consider the \ac{DL} of a single-cell multi-user system, where the \ac{BS} equipped with $\Ntx$ transmit antennas serves $J$ single-antenna \acp{MT}.
We adopt linear precoding such that the precoded \ac{DL} data vector is given as $\mbx = \sum_{j=1}^{J}\mbv_js_j$, where $s_j$ is the transmit signal of \ac{MT} $j$, with $\expec[s_j]=0$ and $\expec[\left|s_j\right|^2]=1$, and $\mbv_j \in \C^{\Ntx}$ is the precoding vector of \ac{MT} $j$.
The precoders satisfy the transmit power constraint $\sum_{j=1}^J \operatorname{tr}(\mbv_j^\herm \mbv_j) = \rho$. 
The sum-rate is given as
\begin{equation} 
\label{eq:inst_sumrate}
    R = \sum_{j=1}^J \log_2 \Bigg(1 + \dfrac{\left|\mbh_j^\tp\mbv_j\right|^2}{ \sum_{m\neq j} \left|\mbh_j^\tp\mbv_m\right|^2 + \sigma_n^2}\Bigg),
\end{equation}
where $\mbh_j\in \C^{\Ntx}$ denotes the channel of \ac{MT} $j$ and $\sigma_n^2$ is the noise variance.
In the considered limited feedback setup, the task of the \ac{BS} is to design the precoders $\mbv_j$ given each \ac{MT}'s feedback information~$k^\star_j$, which is encoded by $B$ bits.

\subsection{Pilot Transmission Phase}

Prior to data transmission, $n_p$ commonly known orthogonal pilots are broadcasted to all \acp{MT}, from which each \ac{MT} infers its feedback information~$k^\star_j$ (see \eqref{eq:dft_idx}, \eqref{eq:ecsi_index_j}, or \eqref{eq:pcsi_index}).
The received signal of each \ac{MT} in the pilot transmission phase is given by
\begin{equation} \label{eq:noisy_obs}
    \mby_j = \mbP \mbh_j + \mbn_j \in \C^{n_p}
\end{equation}
with the \ac{AWGN} $\mbn_j \sim \mathcal{N}_\C(\mathbf{0}, \mbSigma)$ with $\mbSigma = \sigma_n^2 \mathbf{I}_{n_p}$.
We use a $2$D-DFT (sub)matrix, which is obtained through the Kronecker product of two \ac{DFT} matrices, as the pilot matrix since a \ac{URA} is employed at the \ac{BS}, see, e.g., \cite{TsZhWa18}.
To satisfy the power constraint, we normalize each column $\mbp_\ell$ of $\mbP^\tp$, for all $\ell \in \{1,2, \dots, n_p\} $, i.e., $\|\mbp_\ell\|_2^2=\rho$.
We examine scenarios with fewer pilots than transmit antennas, i.e., $n_p < \Ntx$.

\subsection{Real-World Channel Data} \label{sec:data_generation}

\begin{figure}
    \centering
    \begin{tikzpicture}
        \draw (0,0) node[below right] {\includegraphics[width=0.975\columnwidth]{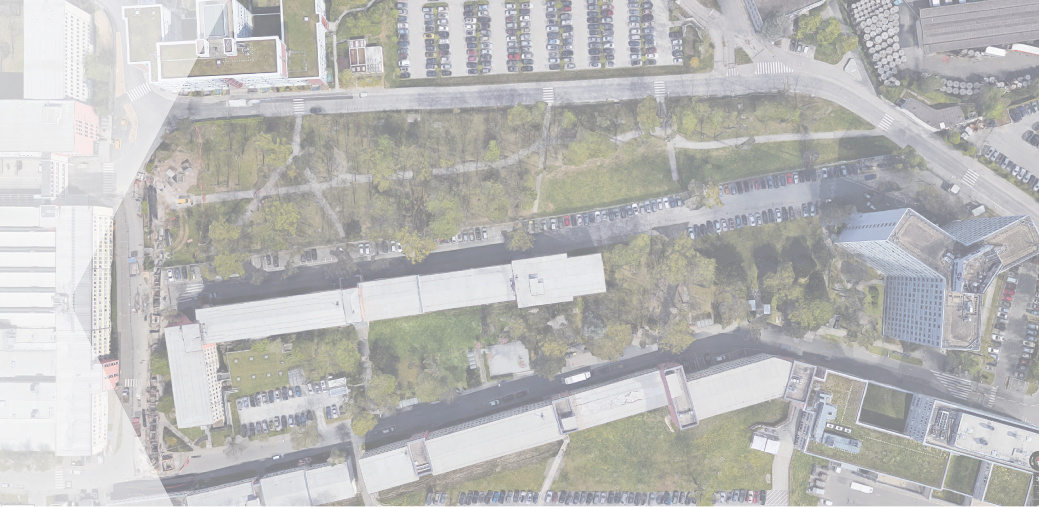}};
        \draw (0,0) node[below right] {\includegraphics[width=0.975\columnwidth]{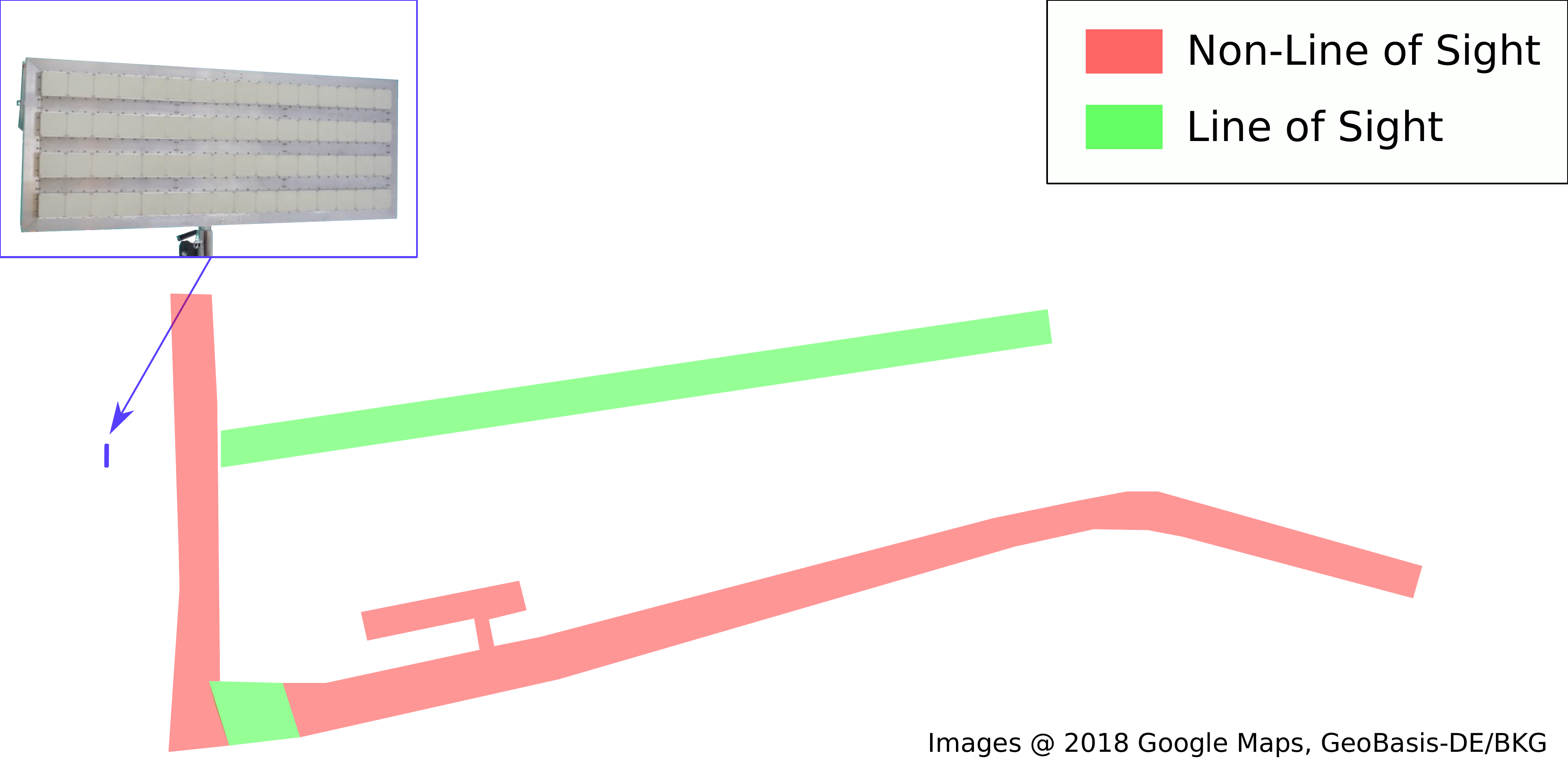}};
    \end{tikzpicture}
    \caption{Nokia campus in Stuttgart, Germany---Measurement environment.}
    \label{fig:meas_campaign}
\end{figure}

The measurement campaign was carried out at the Nokia campus in Stuttgart, Germany, in October/November 2017.
As illustrated in \Cref{fig:meas_campaign}, the \ac{BS} antenna with a down-tilt of $\SI{10}{\degree}$ was installed on a rooftop approximately $\SI{20}{m}$ above the ground.
It comprises a \ac{URA} with $\Nv=4$ vertical and $\Nh=16$ horizontal single polarized patch antennas, yielding in total $N=64$ antenna elements.
The \ac{BS} antenna array geometry was tailored to the \ac{UMi} propagation scenario, characterized by a larger horizontal than vertical angular spread.
Thus, the horizontal spacing was set to $\lambda/2$, and the vertical spacing was~$\lambda$, where $\lambda$ is the wavelength. 
The carrier frequency employed was $\SI{2.18}{\giga\hertz}$. 
The single monopole receive antenna, emulating the \acp{MT}, was attached atop a mobile vehicle at a height of $\SI{1.5}{m}$. 
The vehicle's maximum speed during the campaign was $\SI{25}{kmph}$. 
Further details about the measurement campaign are available in \cite{HeDeWeKoUt19, TuFeGrKoUt22}.

\section{Conventional DFT Codebook-based Feedback} \label{sec:conv_mumimo}

In conventional codebook-based feedback schemes, each \ac{MT} first calculates an estimate of its channel $\hat{\mbh}_j$ and then determines the feedback information as (see, e.g.,~\cite{KaKoGeKn09}):
\begin{equation} \label{eq:dft_idx}
    k^\star_j = \argmax_{k } |\mbc_k^\herm \hat{\mbh}_j|
\end{equation}
where $\mbc_k \in \mathcal{C}$, with $\mathcal{C} = \{\mbc_1, \dots, \mbc_K \}$ being a codebook of cardinality $|\mathcal{C}| = K = 2^B$, with $B$ bits.
In scenarios with spatial correlation due to specific antenna array geometries at the \ac{BS}, \ac{DFT}-based codebooks offer well-established solutions \cite{YaYaHa10, LiSuZeZhYuXiXu13}. 
Since we have a \ac{URA} deployed at the \ac{BS} side, a 2D-\ac{DFT} codebook, which is constructed by the Kronecker product of two under-/oversampled \ac{DFT} matrices $\mbF_{\Nv} \in \C^{\Nv \times S_{\mathrm{v}}\Nv}$ and $\mbF_{\Nh} \in \C^{\Nh \times S_{\mathrm{h}}\Nh}$, i.e., the codebook entries $\mbc_i \in \mathcal{C}$, with normalization $\|\mbc_i\|_2=1$, are given by the columns of the matrix $\mbF_{\Nv} \otimes \mbF_{\Nh}$~\cite{LiSuZeZhYuXiXu13}.
The under-/oversampling factors $S_{\mathrm{v}}$ and $S_{\mathrm{h}}$ are chosen to satisfy $|\mathcal{C}| = K = 2^B$ for a given value of $B$, accordingly.
Given the feedback $k^\star_j$ of each \ac{MT}, the \ac{BS} represents the channel of each \ac{MT} as \cite{Ji06},
\begin{equation} \label{eq:dft_approach}
    \tilde{\mbh}_j = \mbc_{k^\star_j},
\end{equation}
and utilizes the \ac{RCI} method from \cite{PeHoSw05}, with regularization factor $\frac{J \sigma_n^2}{\rho}$, in order to jointly design the precoders.

\section{GMM-based Feedback Scheme} \label{sec:gmm_based_scheme}

Any channel $\mbh$ of the environment depicted in \Cref{fig:meas_campaign} follows the \ac{PDF} $f_{\mbh}$.
Particularly, the channels $\mbh_j$ of \acp{MT} located anywhere within the \ac{BS}'s coverage area are realizations of a random variable with \ac{PDF} $f_{\mbh}$.
This \ac{PDF} is not available analytically.
The scheme proposed in \cite{TuFeKoJoUt23} utilizes a \ac{GMM} to approximate this \ac{PDF} and thereby captures the environment knowledge.
The learned model is then subsequently transferred to the \acp{MT} to equip them with model awareness.
In the online phase, the \acp{MT} can exploit this knowledge to infer their feedback information.
After the \acp{MT} report their feedback to the \ac{BS}, it designs the precoders utilizing either a directional information-based method or a generative modeling-based approach.

\subsection{Capturing the Environment at the BS -- Offline}

The characteristics of the environment are captured offline with a \ac{GMM}, which consists of $K=2^B$ components:
\begin{equation}\label{eq:gmm_of_h}
    f^{(K)}_{\mbh}(\mbh_j) = \sum\nolimits_{k=1}^K \pi_k \calN_{\C}(\mbh_j; \mbmu_k, \mbC_k).
\end{equation}
Each \ac{GMM}-component is described by the mixing coefficients $\pi_k$, the means $\mbmu_k$, and the covariance matrices $\mbC_k$.
Note that the parameters of the \ac{GMM}, i.e., $\{\pi_k,\mbmu_k, \mbC_k\}_{k=1}^K$, are common to all \acp{MT}.
Given a training dataset \(\mathcal{H} \) (see \Cref{sec:sim_results}), an \ac{EM} algorithm can be used to compute maximum likelihood estimates of the \ac{GMM} parameters, see~\cite[Subsec.~9.2.2]{bookBi06}.

With a \ac{GMM}, the posterior probability that the channel of \ac{MT} $j$ originates from component $k$ can be computed as~\cite[Sec.~9.2]{bookBi06},
\begin{equation}\label{eq:responsibilities_h}
    p(k \mid \mbh_j) = \frac{\pi_k \calN_{\C}(\mbh_j; \mbmu_k, \mbC_k)}{\sum_{i=1}^K \pi_i \calN_{\C}(\mbh_j; \mbmu_i, \mbC_i) }.
\end{equation}
These posterior probabilities are also referred to as responsibilities.


\subsection{Model Transfer to the MTs -- Offline}

To enable the \acp{MT} to infer their feedback information, the knowledge of the \ac{GMM}'s parameters is a prerequisite.
Conceptually, this requires transferring the model parameters, i.e., $\{\pi_k,\mbmu_k, \mbC_k\}_{k=1}^K$, to the \acp{MT} upon entering the coverage area of the \ac{BS}.
This transfer equips the \acp{MT} with environmental awareness and is required only once since the \ac{GMM} is fixed for a specific \ac{BS} environment.

By incorporating model-based insights, the number of \ac{GMM} parameters and, thus, the transfer overhead can be reduced tremendously.
In particular, the \ac{GMM} covariances can be constrained to exhibit a certain structure.
Since we have a \ac{URA} deployed at the \ac{BS}, a reasonable choice is to constrain the \ac{GMM} covariance matrices to be block-Toeplitz matrices with Toeplitz blocks, which are expressed as $\mbC_k = \mbD^\herm \diag(\mbc_{k}) \mbD$, with  $\mbD = \mbD_{\Nv} \otimes \mbD_{\Nh}$, where $\mbD_T$ (with $T \in \{\Nv, \Nh\}$) contains the first $T$ columns of a $2T\times 2T$ \ac{DFT} matrix and $\mbc_k \in \R_{+}^{4N}$, cf.~\cite{TuFeGrKoUt22, TuFeUt23}.
Consequently, the covariance matrices of each component are fully characterized by the vectors $\mbc_{k}$.
In Table \ref{tab:num_params}, we depict the number of covariance parameters for different numbers $B$ of feedback bits.
We can see that the structural constraints decrease the number of covariance parameters.
Further advantages of introducing structural constraints include lower offline training complexities and a reduction of the number of required training samples.

\begin{table}[t]
\renewcommand{\arraystretch}{1.5}
\begin{center}
\begin{tabular}{|l|c|c|c|c|c|c}
\hline
\textbf{Name} & \textbf{Cov. Parameters} & $B=4$ & $B=6$ & $B=8$\\ \hline
Full & \( \frac{1}{2} K N(N+1) \) & \( 3.3 \cdot 10^4 \) & \( 1.3 \cdot 10^5 \) & \( 5.3 \cdot 10^5 \) \\ \hline
Toeplitz & $ 4 K N$ & \( 4.1 \cdot 10^3 \) & \( 1.6 \cdot 10^4 \) & \( 6.6 \cdot 10^4 \) \\ \hline
\end{tabular}
\end{center}
\caption{Model transfer overhead for different numbers $B$ of feedback bits (where $K=2^B$, and $N=64$).}
\label{tab:num_params}
\end{table}

\subsection{Inferring the Feedback Information  at the MTs -- Online}

In the online phase, each \ac{MT}'s task is to infer its feedback information given the pilot observation $\mby_j$ with the help of the \ac{GMM}.
Due to the joint Gaussianity of each \ac{GMM} component (cf. \eqref{eq:gmm_of_h}) together with the \ac{AWGN}, the \ac{GMM} of the observations can be simply computed with the \ac{GMM} from \eqref{eq:gmm_of_h} as
\begin{equation}\label{eq:gmm_y}
    f_{\mby}^{(K)}(\mby_j) = \sum\nolimits_{k=1}^K \pi_k \calN_{\C}(\mby_j; \mbP \meanhk, \mbP \covhk \mbP^\herm + \mbSigma).
\end{equation}

Accordingly, each \ac{MT} $j$ can compute the responsibilities given the observations $\mby_j$ as 
\begin{equation}\label{eq:responsibilities}
    p(k \mid \mby_j) = \frac{\pi_k \calN_{\C}(\mby_j; \mbP \meanhk, \mbP \covhk \mbP^\herm + \mbSigma)}{\sum_{i=1}^K \pi_i \calN_{\C}(\mby_j; \mbP \meanhi, \mbP \covhi \mbP^\herm + \mbSigma) }.
\end{equation}
The feedback information $k_j^\star$ is then determined as
\begin{equation} \label{eq:ecsi_index_j}
    k^\star_j = \argmax_{k } ~{p(k \mid \mby_j)},
\end{equation}
i.e., the index with the highest responsibility of the observed pilot signal $\mby_j$ of each \ac{MT} serves as the respective feedback information.
The responsibilities evaluate how well each component of the \ac{GMM} explains the underlying channel $\mbh_j$ of the observed pilot signal $\mby_j$.
The feedback information is thus simply the index of the \ac{GMM} component that explains the channel best.

As a reference for performance analysis, perfect \ac{CSI} can be used to determine the feedback information 
\begin{equation} \label{eq:pcsi_index}
    k^\star_j = \argmax_{k } ~{p(k \mid \mbh_j)}.
\end{equation}

\subsection{Designing the Precoders at the \ac{BS}  -- Online}

Given the feedback information $k_j^\star$ of each \ac{MT}, the \ac{BS} has two options for jointly designing the precoders $\mbv_j$.

\subsubsection{Directional Information-based Precoder Design} \label{sec:dirinfo}

The first option utilizes the directional information, which can be associated with each \ac{GMM} component, in order to approximate each \ac{MT}'s channel:
\begin{equation} \label{eq:dir_approach}
    \tilde{\mbh}_j = \text{eigv}(\mbC_{k_j^\star} + \mbmu_{k_j^\star}
    \mbmu_{k_j^\star}^\herm),
\end{equation}
where $\text{eigv}(\mbX)$ extracts the dominant eigenvector of $\mbX$.
Accordingly, the channel is represented by the dominant eigenvector of the correlation matrix of the respective \ac{GMM} component.
Note that using the dominant eigenvector of the correlation matrix as representative is inspired by the centroid condition of the Lloyd clustering approach, cf., e.g., \cite{PeGi06}.
The dominant eigenvectors of the correlation matrices per \ac{GMM} component can be precomputed in the offline phase.
With the representations of the \acp{MT}' channels as found in \eqref{eq:dir_approach}, the \ac{BS} can then design the precoders employing the \ac{RCI} method, see~\cite{PeHoSw05}.

With the \ac{GMM}-based feedback approach, an iterative update of the codebook, as is the case for Lloyd's clustering approach, is circumvented.
In principle, one would have to partition the training data utilizing the \ac{GMM} into $K$ disjoint sets as proposed in \cite{TuFeKoJoUt23} in order to calculate cluster representatives, cf.~\cite{TuFeKoJoUt23}. 
In the case of single-antenna \acp{MT}, the cluster representatives are computed in closed form via~\eqref{eq:dir_approach}.

\subsubsection{Generative Modeling-based Precoder Design} \label{sec:genmod}

The second option is based on generative modeling, where the channels of the \acp{MT} are treated as random variables, and a stochastic version of the well-known iterative \ac{WMMSE} algorithm~\cite{ShRaLuHe11}, i.e., the \ac{SWMMSE} algorithm from \cite{RaBoLu13, RaSaLu16} is employed for precoder design.
In particular, with the feedback information $k_j^\star$ of each \ac{MT}, see \eqref{eq:ecsi_index_j}, the \ac{GMM} enables the generation of samples via
\begin{equation}
     \mbh_{j,\text{sample}} \sim \calN_{\C}(\mbmu_{k^\star_j}, \mbC_{k^\star_j}),
\end{equation}
which resemble the distribution of the channel of \ac{MT} $j$.
The \ac{BS} utilizes these generated samples in each iteration step of the \ac{SWMMSE} algorithm in order to jointly design the precoders, cf. \cite{TuFeKoJoUt23}.

\subsection{Complexity Analysis} \label{sec:comp_ana}

The online computational complexity for inferring the feedback information of the \ac{GMM}-based feedback scheme is dominated by matrix-vector multiplications.
This is because the computation of the responsibilities in~\eqref{eq:responsibilities} involves evaluating Gaussian densities, and the associated determinant and inverse expressions can be pre-calculated for a specific \ac{SNR} level due to the fixed \ac{GMM} parameters.
Accordingly, inferring the feedback information at the \acp{MT} via \eqref{eq:ecsi_index_j} in the online phase has a complexity of \( \calO(K n_p^2) \).
A significant advantage of the \ac{GMM}-based feedback scheme is that the complexity is independent of the number $\Ntx$ of transmit antennas.
This is especially beneficial for massive \ac{MIMO} systems where the \ac{BS} is typically equipped with a large antenna array.
Additionally, the \ac{GMM}-based feedback scheme enables parallelization concerning the number of components $K$, allowing the simultaneous evaluation of all $K$ responsibilities.


\section{Discussion: \\Generative Modeling versus DFT Codebooks}

Although \ac{DFT}-based codebook solutions are well-established in scenarios with spatial correlation, they heavily depend on accurate \ac{DL} \ac{CSI} estimation since the knowledge of the channel is a prerequisite to determine the best-fitting codebook entry. 
This generally holds for codebook-based solutions, including data-based codebook design approaches (cf. \cite{BhSaDh20, PeGi06}), which utilize Lloyd's algorithm.
Obtaining accurate channel estimates at the \acp{MT} often requires the transmission of numerous pilots from the \ac{BS} to the \acp{MT}.
However, in massive \ac{MIMO} systems, where the \ac{BS} typically features many antenna elements, the pilot overhead necessary for full channel illumination is unaffordable \cite{BjLaMa16}.
As a result, there is considerable interest in feedback schemes that perform well even when the number of pilots is fewer than the number of transmit antennas.
The \ac{GMM}-based feedback scheme provides an ideal solution by circumventing explicit \ac{DL} \ac{CSI} estimation and instead directly deducing feedback information from pilot observations.
Intuitively, the \ac{GMM}-based feedback scheme is well-suited for this task because it can effectively handle the presented prior information captured during training and exploit it online to determine the feedback information.

Additionally, the \ac{GMM}-based feedback scheme provides a low-complexity solution for inferring feedback information directly from pilot observations and even supports parallelization, see Subsection \ref{sec:comp_ana}.
This contrasts with the \ac{DFT}-based codebook solution, which undergoes a two-stage process involving channel estimation and subsequent determination of feedback information, introducing additional complexity.

Moreover, as opposed to the \ac{DFT}-based codebook approach, which solely permits the usage of directional information for precoder design, the \ac{GMM}-based feedback scheme additionally enables generative modeling-based precoder design due to the sample generation capability of the \ac{GMM}.
These options allow for a trade-off between improved performance and required processing time for designing the precoders (see \Cref{sec:sim_results}).

Finally, the \ac{GMM} framework even allows the \acp{MT} to estimate their channel by computing an observation-dependent convex combination of component-wise \ac{LMMSE} filters, which are parametrized by the \ac{GMM}, cf. \cite{KoFeTuUt21J}.
The \ac{GMM}-based channel estimator outperforms many other state-of-the-art estimators and thus even enhances the overall performance of the \ac{DFT} codebook-based approach if used before codebook entry selection.
We include this baseline in the simulation results for performance comparison.
Nevertheless, it has to be noted that the computation of a channel estimate via the \ac{GMM} already exhibits higher complexity than the feedback scheme from \Cref{sec:gmm_based_scheme}.

\section{Simulation Results} \label{sec:sim_results}

The \ac{GMM} is fitted using the training set $\mathcal{H}$, which consists of $L =10^5$ samples from the measurement scenario described in Subsection~\ref{sec:data_generation}.
We use another dataset for evaluation purposes, which consists of $10^4$ channels.
The data samples are normalized to satisfy \( \expec[\|\mbh\|^2] = N \). 
Additionally, we fix $\rho=1$, enabling the definition of the \ac{SNR} as \( \frac{1}{\sigma_n^2} \).
We employ the sum-rate as the performance metric, averaging over $500$ multi-user constellations where we randomly select $J$ \acp{MT} from our evaluation set for each constellation.

As outlined in \Cref{sec:conv_mumimo}, the process of the \ac{DFT} codebook entry selection necessitates channel estimation.
In this regard, we consider four different channel estimators, briefly summarized below.
One of these methods is the recently introduced \ac{GMM}-based channel estimator \( \hhat_{\text{GMM}} \) from \cite{KoFeTuUt21J}.
This estimator leverages the same \ac{GMM} obtained through model transfer at the \acp{MT} as detailed in Section~\ref{sec:gmm_based_scheme} and computes a convex combination of per-component \ac{LMMSE} estimates.
Similarly, the \ac{GMM} with a Toeplitz constraint enforced on the covariance matrices can be used and is denoted by \( \hhat_{\text{tGMM}} \).
Another baseline method is the \ac{LMMSE} estimator $\hhat_{\text{LMMSE}}$, where the sample covariance matrix is constructed based on the set $\mathcal{H}$.
Lastly, we employ a compressive sensing estimation technique $\hhat_{\text{OMP}}$ utilizing the \ac{OMP}, cf.~\cite{AlLeHe15}.
Please refer to \cite{TuFeKoJoUt23} for more details.

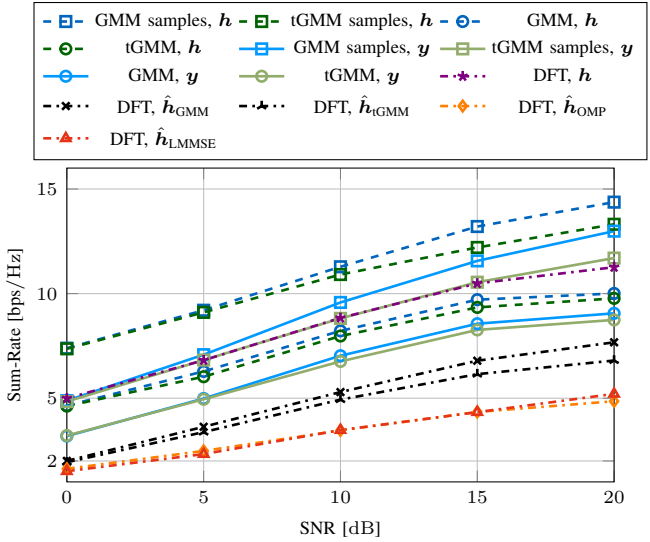
\begin{figure}[tb]
    \centering
        \begin{tikzpicture}
        \begin{axis}[
            height=\largeplotheight,
            width=\largeplotwidth,
            legend pos= north east,
            legend style={font=\scriptsize, at={(0.5,1.025)}, anchor=south, legend columns=3},
            label style={font=\scriptsize},
            tick label style={font=\scriptsize},
            y label style={at={(-0.06,0.5)}},
            title style={align=left, font=\scriptsize},
            xmin=0,
            xmax=20,
            ymin=1,
            ymax=16,
            ytick={2,5,10,15},
            xlabel={SNR [$\SI{}{dB}$]},
            ylabel={Sum-Rate [bps$/\SI{}{\hertz}$]},
            grid=both,
        ]
            \addplot[MUswmmsegmmpgdulperfect, TUMBlue, dashed, mark=square,mark size=2pt]
                table[x=snr,y=SWMMSE_GMM_perfectCSI,col sep=comma] {gmm_measurement/K8_B6/sumrate_over_snr_np8.csv};
                \addlegendentry{GMM samples, $\mbh$};

            \addplot[MUswmmsegmmpgdulperfect, ourdarkgreen, dashed, mark=square,mark size=2pt]
                table[x=snr,y=SWMMSE_GMM_perfectCSI,col sep=comma] {gmm_measurement/K8_B6/sumrate_over_snr_np8_toep.csv};
                \addlegendentry{tGMM samples, $\mbh$};

            \addplot[MUswmmsegmmpgdulperfect, TUMBlue, dashed, mark=o,mark size=2pt]
                table[x=snr,y=RCI_GMM_perfectCSI,col sep=comma] {gmm_measurement/K8_B6/sumrate_over_snr_np8_rciReg.csv};
                \addlegendentry{GMM, $\mbh$};

            \addplot[MUswmmsegmmpgdulperfect, ourdarkgreen, dashed, mark=o,mark size=2pt]
                table[x=snr,y=RCI_GMM_perfectCSI,col sep=comma] {gmm_measurement/K8_B6/sumrate_over_snr_np8_toep_rciReg.csv};
                \addlegendentry{tGMM, $\mbh$};

            \addplot[MUswmmsegmmpgdulfromy, TUMBeamerBlue, solid,mark=square,mark size=2pt]
                table[x=snr,y=SWMMSE_GMM_observation, col sep=comma] {gmm_measurement/K8_B6/sumrate_over_snr_np8.csv};
                \addlegendentry{GMM samples, $\mby$};

            \addplot[MUswmmsegmmpgdulfromy, TUMBeamerGreen, solid,mark=square,mark size=2pt]
                table[x=snr,y=SWMMSE_GMM_observation, col sep=comma] {gmm_measurement/K8_B6/sumrate_over_snr_np8_toep.csv};
                \addlegendentry{tGMM samples, $\mby$};
                
            \addplot[MUswmmsegmmpgdulfromy, TUMBeamerBlue, solid,mark=o,mark size=2pt]
                table[x=snr,y=RCI_GMM_observation, col sep=comma] {gmm_measurement/K8_B6/sumrate_over_snr_np8_rciReg.csv};
                \addlegendentry{GMM, $\mby$};
                
            \addplot[MUswmmsegmmpgdulfromy, TUMBeamerGreen, solid,mark=o,mark size=2pt]
                table[x=snr,y=RCI_GMM_observation, col sep=comma] {gmm_measurement/K8_B6/sumrate_over_snr_np8_toep_rciReg.csv};
                \addlegendentry{tGMM, $\mby$};
                
            \addplot[MUlloydpgdulperfect,dash dot,mark=star,mark size=2pt]
                table[x=snr,y=RCI_DFT_perfectCSI,col sep=comma] {gmm_measurement/K8_B6/sumrate_over_snr_np8_toep_rciReg.csv};
                \addlegendentry{\pltMUdftperfect};
            \addplot[MUlloydpgdulgmmulall,dash dot,mark=x,mark size=2pt]
                table[x=snr,y=RCI_DFT_hGMM,col sep=comma] {gmm_measurement/K8_B6/sumrate_over_snr_np8_rciReg.csv};
                \addlegendentry{\pltMUdftgmmulall};
            \addplot[MUlloydpgdulgmmulall,dash dot,mark=Mercedes star,mark size=2pt]
                table[x=snr,y=RCI_DFT_hGMM,col sep=comma] {gmm_measurement/K8_B6/sumrate_over_snr_np8_toep_rciReg.csv};
                \addlegendentry{DFT, $\hat{\mbh}_{\text{tGMM}}$};
            \addplot[MUlloydpgdulomp,dash dot,mark=diamond,mark size=2pt]
                table[x=snr,y=RCI_DFT_hOMP,col sep=comma] {gmm_measurement/K8_B6/sumrate_over_snr_np8_rciReg.csv};
                \addlegendentry{\pltMUdftomp};
            \addplot[MUlloydpgdulscov,dash dot,mark=triangle,mark size=2pt]
                table[x=snr,y=RCI_DFT_hLMMSE,col sep=comma] {gmm_measurement/K8_B6/sumrate_over_snr_np8_rciReg.csv};
                \addlegendentry{\pltMUdftscov};
        \end{axis}
    \end{tikzpicture}
    \caption{The sum-rate over the \ac{SNR} for a system with $B=6$ feedback bits, $J=8$ \acp{MT}, and $n_p=8$ pilots.}
    \label{fig:meas_K8B6_oversnr}
\end{figure}

For clarity, we exclude the index $j$ in the subsequent descriptions in the legend.
In the following discussion, ``\{GMM, tGMM\}, \{$\mbh, \mby$\},'' denotes the case where either perfect \ac{CSI}, cf.~\eqref{eq:pcsi_index}, is assumed, or the observations $\mby_j$ at each \ac{MT} are utilized, cf.~\eqref{eq:ecsi_index_j}, for determining a feedback index through the \ac{GMM}-based feedback encoding scheme with either full or Toeplitz-structured covariances.
Subsequently, the channel of each \ac{MT} is represented by the directional information associated with the respective \ac{GMM} component, cf. Subsection~\ref{sec:dirinfo}.
With ``DFT, \{$\mbh$, $\hhat_{\text{GMM}}$, $\hhat_{\text{tGMM}}$, $\hhat_{\text{OMP}}, \hhat_{\text{LMMSE}}$\},'' we denote cases where either perfect \ac{CSI} is used or the channel is estimated at each \ac{MT}.
Afterward, the feedback information is determined using the DFT codebook, cf.~\Cref{sec:conv_mumimo}.
These methods use the \ac{RCI} method for precoder design.
Lastly, with ``\{GMM, tGMM\} samples, \{$\mbh, \mby$\},'' we refer to the generative modeling-based precoder design approach utilizing the \ac{SWMMSE} algorithm, cf. Subsection~\ref{sec:genmod}, with the maximum number of iterations set to $I_{\max}=300$.

In \Cref{fig:meas_K8B6_oversnr}, we set $B=6$ bits, $J=8$ \acp{MT}, and $n_p=8$ pilots and depict the sum-rate over the \ac{SNR}.
In the case of perfect \ac{CSI} assumed at each \ac{MT}, we can observe that ``GMM samples, $\mbh$'' performs best.
The benefit of reducing the model transfer overhead realized by the structural constraint enforced on the \ac{GMM} covariances comes at a cost of slightly degraded performance, i.e., ``tGMM samples, $\mbh$'' performs slightly worse.
Both approaches clearly outperform the \ac{DFT}-based codebook approach ``DFT, $\mbh$''.
Remarkably, the generative modeling-based precoder design approaches using solely a reduced number of pilot observations ``\{GMM, tGMM\} samples, $\mby$'' outperform or yield a similar performance as ``DFT, $\mbh$''.
This highlights the substantial potential offered by the sample generation capability of the \ac{GMM}.
The ``DFT, $\mbh$'' approach, in turn, outperforms the approaches ``\{GMM, tGMM\}, $\mbh$,'' which utilize the directional information associated with each \ac{GMM} component.

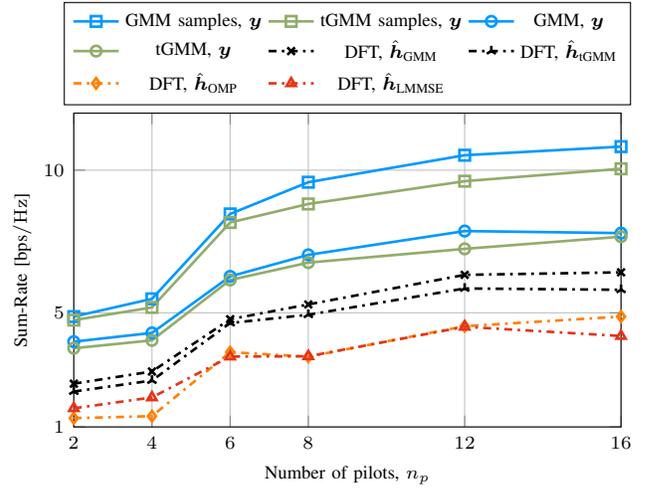
\begin{figure}[tb]
    \centering
        \begin{tikzpicture}
        \begin{axis}[
            height=\largeplotheight,
            width=\largeplotwidth,
            legend pos= north east,
            legend style={font=\scriptsize, at={(0.5,1.025)}, anchor=south, legend columns=3},
            label style={font=\scriptsize},
            tick label style={font=\scriptsize},
            y label style={at={(-0.06,0.5)}},
            title style={align=left, font=\scriptsize},
            xmin=2,
            xmax=16,
            xtick={2,4,6,8,12, 16},
            ymin=1,
            ymax=12,
            ytick={1,5,10,15},
            xlabel={Number of pilots, $n_p$},
            ylabel={Sum-Rate [bps$/\SI{}{\hertz}$]},
            grid=both,
        ]
            \addplot[MUswmmsegmmpgdulfromy, TUMBeamerBlue, solid,mark=square,mark size=2pt]
                table[x=pilots,y=SWMMSE_GMM_observation, col sep=comma] {gmm_measurement/K8_B6/sumrate_over_pilots_snr10.csv};
                \addlegendentry{GMM samples, $\mby$};
                

            \addplot[MUswmmsegmmpgdulfromy, TUMBeamerGreen, solid,mark=square,mark size=2pt]
                table[x=pilots,y=SWMMSE_GMM_observation, col sep=comma] {gmm_measurement/K8_B6/sumrate_over_pilots_snr10_toep.csv};
                \addlegendentry{tGMM samples, $\mby$};
                

            \addplot[MUswmmsegmmpgdulfromy, TUMBeamerBlue, solid,mark=o,mark size=2pt]
                table[x=pilots,y=RCI_GMM_observation, col sep=comma] {gmm_measurement/K8_B6/sumrate_over_pilots_snr10_rciReg.csv};
                \addlegendentry{GMM, $\mby$};
                

            \addplot[MUswmmsegmmpgdulfromy, TUMBeamerGreen, solid,mark=o,mark size=2pt]
                table[x=pilots,y=RCI_GMM_observation, col sep=comma] {gmm_measurement/K8_B6/sumrate_over_pilots_snr10_toep_rciReg.csv};
                \addlegendentry{tGMM, $\mby$};
                

            \addplot[MUlloydpgdulgmmulall,dash dot,mark=x,mark size=2pt]
                table[x=pilots,y=RCI_DFT_hGMM,col sep=comma] {gmm_measurement/K8_B6/sumrate_over_pilots_snr10_rciReg.csv};
                \addlegendentry{\pltMUdftgmmulall};
            \addplot[MUlloydpgdulgmmulall,dash dot,mark=Mercedes star,mark size=2pt]
                table[x=pilots,y=RCI_DFT_hGMM,col sep=comma] {gmm_measurement/K8_B6/sumrate_over_pilots_snr10_toep_rciReg.csv};
                \addlegendentry{DFT, $\hat{\mbh}_{\text{tGMM}}$};
            \addplot[MUlloydpgdulomp,dash dot,mark=diamond,mark size=2pt]
                table[x=pilots,y=RCI_DFT_hOMP,col sep=comma] {gmm_measurement/K8_B6/sumrate_over_pilots_snr10_rciReg.csv};
                \addlegendentry{\pltMUdftomp};
            \addplot[MUlloydpgdulscov,dash dot,mark=triangle,mark size=2pt]
                table[x=pilots,y=RCI_DFT_hLMMSE,col sep=comma] {gmm_measurement/K8_B6/sumrate_over_pilots_snr10_toep_rciReg.csv};
                \addlegendentry{\pltMUdftscov};
        \end{axis}
    \end{tikzpicture}
    \caption{The sum-rate over the number $n_p$ of pilots for a system with $J=8$ \acp{MT}, and $\text{SNR}=\SI{10}{dB}$.}
    \label{fig:meas_K8B6_overpilots}
\end{figure}

However, the imperfect channel knowledge at the \acp{MT} due to channel estimation errors tremendously degrades the performance of the \ac{DFT}-based codebook approaches ``DFT, \{$\hhat_{\text{GMM}}$, $\hhat_{\text{tGMM}}$, $\hhat_{\text{OMP}}, \hhat_{\text{LMMSE}}$\}''.
The \ac{GMM}-based approaches using the pilot observations ``\{GMM, tGMM\}, $\mby$'' exhibit higher robustness, and as a consequence outperform ``DFT, \{$\hhat_{\text{GMM}}$, $\hhat_{\text{tGMM}}$, $\hhat_{\text{OMP}}, \hhat_{\text{LMMSE}}$\},'' irrespective of the used channel estimator.
Utilizing the \ac{GMM}-based channel estimators $\hhat_{\text{GMM}}$ and $\hhat_{\text{tGMM}}$ before codebook entry selection with the \ac{DFT}-based codebook approach is superior as compared to using the other channel estimators.
The loss in performance due to structural constraints can be observed similarly in this case.

However, assuming the availability of perfect \ac{CSI} at the \ac{MT} in the online phase is unrealistic.
In the remainder, we focus our analysis on systems characterized by low pilot overhead $(n_p < N)$.
In \Cref{fig:meas_K8B6_overpilots}, we assess the impact of the number of pilots $n_p$ on the performance whereby $B=6$ bits, $J=8$ \acp{MT} for a fixed $\text{SNR}=\SI{10}{dB}$.
The directional and the generative modeling-based approaches utilizing the \ac{GMM} outperform the \ac{DFT}-based codebook approach for all considered numbers of pilots $n_p$ by a large margin.
We can see that ``\{GMM, tGMM\} samples, $\mby$'' or ``\{GMM, tGMM\}, $\mby$'' only require $n_p=6$ pilots to achieve the same or better performance as ``DFT, \{$\hhat_{\text{GMM}}$, $\hhat_{\text{tGMM}}$, $\hhat_{\text{OMP}}, \hhat_{\text{LMMSE}}$\},'' which require $n_p=16$ pilots.
Accordingly, systems with lower pilot overhead can be deployed without sacrificing performance due to this robustness of the \ac{GMM}-based feedback scheme.

\begin{figure}[tb]
    \centering
        \begin{tikzpicture}
        \begin{axis}[
            height=\largeplotheight,
            width=\largeplotwidth,
            legend pos= north east,
            legend style={font=\scriptsize, at={(0.5,1.025)}, anchor=south, legend columns=3},
            label style={font=\scriptsize},
            tick label style={font=\scriptsize},
            y label style={at={(-0.06,0.5)}},
            title style={align=left, font=\scriptsize},
            xmin=4,
            xmax=8,
            xtick={4,5,6,7,8},
            ymin=1,
            ymax=13,
            ytick={1,3,5,10,15},
            xlabel={Feedback Bits, $B$},
            ylabel={Sum-Rate [bps$/\SI{}{\hertz}$]},
            grid=both,
        ]
            \addplot[MUswmmsegmmpgdulfromy, TUMBeamerBlue, solid,mark=square,mark size=2pt]
                table[x=bits,y=SWMMSE_GMM_observation, col sep=comma] {gmm_measurement/K8_B6/sumrate_over_bits_snr10_np8.csv};
                \addlegendentry{GMM samples, $\mby$};
                

            \addplot[MUswmmsegmmpgdulfromy, TUMBeamerGreen, solid,mark=square,mark size=2pt]
                table[x=bits,y=SWMMSE_GMM_observation, col sep=comma] {gmm_measurement/K8_B6/sumrate_over_bits_snr10_toep_np8_rciReg.csv};
                \addlegendentry{tGMM samples, $\mby$};
                

            \addplot[MUswmmsegmmpgdulfromy, TUMBeamerBlue, solid,mark=o,mark size=2pt]
                table[x=bits,y=RCI_GMM_observation, col sep=comma] {gmm_measurement/K8_B6/sumrate_over_bits_snr10_np8_rciReg.csv};
                \addlegendentry{GMM, $\mby$};
                

            \addplot[MUswmmsegmmpgdulfromy, TUMBeamerGreen, solid,mark=o,mark size=2pt]
                table[x=bits,y=RCI_GMM_observation, col sep=comma] {gmm_measurement/K8_B6/sumrate_over_bits_snr10_toep_np8_rciReg.csv};
                \addlegendentry{tGMM, $\mby$};
                

            \addplot[MUlloydpgdulgmmulall,dash dot,mark=x,mark size=2pt]
                table[x=bits,y=RCI_DFT_hGMM,col sep=comma] {gmm_measurement/K8_B6/sumrate_over_bits_snr10_np8_rciReg.csv};
                \addlegendentry{\pltMUdftgmmulall};
            \addplot[MUlloydpgdulgmmulall,dash dot,mark=Mercedes star,mark size=2pt]
                table[x=bits,y=RCI_DFT_hGMM,col sep=comma] {gmm_measurement/K8_B6/sumrate_over_bits_snr10_toep_np8_rciReg.csv};
                \addlegendentry{DFT, $\hat{\mbh}_{\text{tGMM}}$};
            \addplot[MUlloydpgdulomp,dash dot,mark=diamond,mark size=2pt]
                table[x=bits,y=RCI_DFT_hOMP,col sep=comma] {gmm_measurement/K8_B6/sumrate_over_bits_snr10_np8_rciReg.csv};
                \addlegendentry{\pltMUdftomp};
            \addplot[MUlloydpgdulscov,dash dot,mark=triangle,mark size=2pt]
                table[x=bits,y=RCI_DFT_hLMMSE,col sep=comma] {gmm_measurement/K8_B6/sumrate_over_bits_snr10_np8_rciReg.csv};
                \addlegendentry{\pltMUdftscov};
        \end{axis}
    \end{tikzpicture}
    \caption{The sum-rate over the number $B$ of feedback bits for a system with $J=8$ \acp{MT}, $n_p=8$ pilots, and $\text{SNR}=\SI{10}{dB}$.}
    \label{fig:meas_K8B6_overbits}
\end{figure}
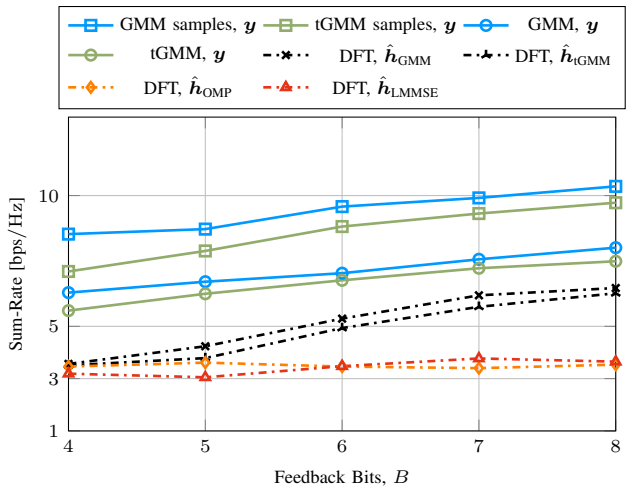

In \Cref{fig:meas_K8B6_overbits}, we fix $J=8$ \acp{MT}, $n_p=8$ pilots, and $\text{SNR}=\SI{10}{dB}$ and investigate the impact of the number $B$ of feedback bits on the system performance.
We can observe that again, the \ac{GMM}-based approaches  ``\{GMM, tGMM\} samples, $\mby$'' and ``\{GMM, tGMM\}, $\mby$'' perform best.
The results suggest that systems with fewer bits can be used to achieve the same performance as with the \ac{DFT}-based codebook approaches  ``DFT, \{$\hhat_{\text{GMM}}$, $\hhat_{\text{tGMM}}$, $\hhat_{\text{OMP}}, \hhat_{\text{LMMSE}}$\},'' which would decrease not only the required feedback overhead but also the required processing for inferring the feedback information at the \acp{MT}.

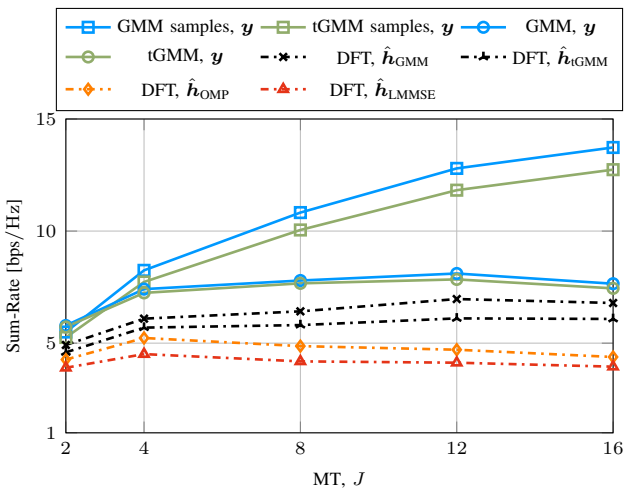
\begin{figure}[tb]
    \centering
        \begin{tikzpicture}
        \begin{axis}[
            height=\largeplotheight,
            width=\largeplotwidth,
            legend pos= north east,
            legend style={font=\scriptsize, at={(0.5,1.025)}, anchor=south, legend columns=3},
            label style={font=\scriptsize},
            tick label style={font=\scriptsize},
            y label style={at={(-0.06,0.5)}},
            title style={align=left, font=\scriptsize},
            xmin=2,
            xmax=16,
            xtick={2,4,8,12,16},
            ymin=1,
            ymax=15,
            ytick={1,5,10,15},
            xlabel={\ac{MT}, $J$},
            ylabel={Sum-Rate [bps$/\SI{}{\hertz}$]},
            grid=both,
        ]
            \addplot[MUswmmsegmmpgdulfromy, TUMBeamerBlue, solid,mark=square,mark size=2pt]
                table[x=user,y=SWMMSE_GMM_observation, col sep=comma] {gmm_measurement/K8_B6/sumrate_over_user_snr10_np16.csv};
                \addlegendentry{GMM samples, $\mby$};
                

            \addplot[MUswmmsegmmpgdulfromy, TUMBeamerGreen, solid,mark=square,mark size=2pt]
                table[x=user,y=SWMMSE_GMM_observation, col sep=comma] {gmm_measurement/K8_B6/sumrate_over_user_snr10_toep_np16.csv};
                \addlegendentry{tGMM samples, $\mby$};
                

            \addplot[MUswmmsegmmpgdulfromy, TUMBeamerBlue, solid,mark=o,mark size=2pt]
                table[x=user,y=RCI_GMM_observation, col sep=comma] {gmm_measurement/K8_B6/sumrate_over_user_snr10_np16_rciReg.csv};
                \addlegendentry{GMM, $\mby$};
                

            \addplot[MUswmmsegmmpgdulfromy, TUMBeamerGreen, solid,mark=o,mark size=2pt]
                table[x=user,y=RCI_GMM_observation, col sep=comma] {gmm_measurement/K8_B6/sumrate_over_user_snr10_toep_np16.csv};
                \addlegendentry{tGMM, $\mby$};
                

            \addplot[MUlloydpgdulgmmulall,dash dot,mark=x,mark size=2pt]
                table[x=user,y=RCI_DFT_hGMM,col sep=comma] {gmm_measurement/K8_B6/sumrate_over_user_snr10_np16_rciReg.csv};
                \addlegendentry{\pltMUdftgmmulall};
            \addplot[MUlloydpgdulgmmulall,dash dot,mark=Mercedes star,mark size=2pt]
                table[x=user,y=RCI_DFT_hGMM,col sep=comma] {gmm_measurement/K8_B6/sumrate_over_user_snr10_toep_np16.csv};
                \addlegendentry{DFT, $\hat{\mbh}_{\text{tGMM}}$};
            \addplot[MUlloydpgdulomp,dash dot,mark=diamond,mark size=2pt]
                table[x=user,y=RCI_DFT_hOMP,col sep=comma] {gmm_measurement/K8_B6/sumrate_over_user_snr10_np16_rciReg.csv};
                \addlegendentry{\pltMUdftomp};
            \addplot[MUlloydpgdulscov,dash dot,mark=triangle,mark size=2pt]
                table[x=user,y=RCI_DFT_hLMMSE,col sep=comma] {gmm_measurement/K8_B6/sumrate_over_user_snr10_np16_rciReg.csv};
                \addlegendentry{\pltMUdftscov};
        \end{axis}
    \end{tikzpicture}
    \caption{The sum-rate over the number $J$ of \acp{MT} for a system with $B=6$ feedback bits, $n_p=16$ pilots, and $\text{SNR}=\SI{10}{dB}$.}
    \label{fig:meas_KXB6_overuser}
\end{figure}

In \Cref{fig:meas_KXB6_overuser}, we fix $B=6$ bits, $n_p=16$ pilots, and $\text{SNR}=\SI{10}{dB}$ and vary the number $J$ of served \acp{MT}.
It can be seen that the \ac{GMM}-based feedback scheme is superior as compared to the \ac{DFT}-based codebook approaches for all numbers of \acp{MT}.
Remarkably, the sum-rates of the generative modeling-based approaches ``\{GMM, tGMM\} samples, $\mby$'' steadily increase with an increasing number of \acp{MT}.
The remaining approaches quickly saturate or even slightly degrade with an increasing number $J$ of \acp{MT}.
This can be reasoned by the fact that with an increasing number of \acp{MT}, resolving the interference present in the scenario becomes more difficult, particularly when restricted to a codebook of finite size.
In contrast, with the generative modeling-based approach, due to the involved sampling procedure, a different representative interference scenario is provided to and exploited by the \ac{SWMMSE} to design the precoders.

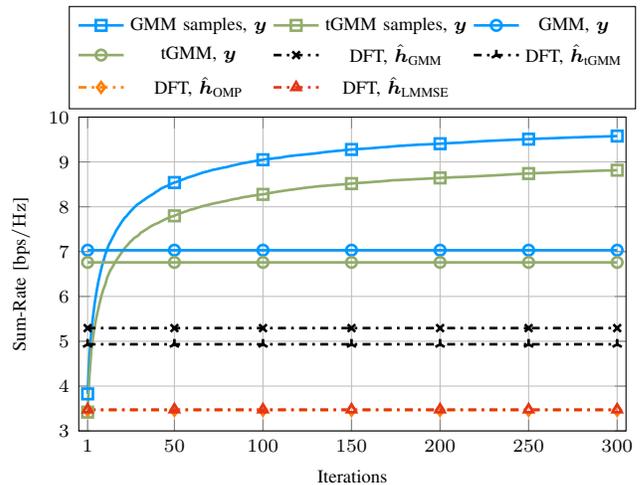
\begin{figure}[tb]
    \centering
        \begin{tikzpicture}[spy using outlines={magnification=2, connect spies}]
        \begin{axis}[
            height=\largeplotheight,
            width=\normalplotwidth,
            legend pos= north east,
            legend style={font=\scriptsize, at={(0.5,1.025)}, anchor=south, legend columns=3},
            legend image post style={mark indices={}},
            label style={font=\scriptsize},
            tick label style={font=\scriptsize},
            title style={align=left, font=\scriptsize},
            xmin=-5,
            xmax=304,
            xtick={0,49,99,149,199,249,299},
            xticklabels={$1$,$50$,$100$,$150$,$200$,$250$,$300$},
            ymin=3,
            ymax=10,
            ytick={3,4,5,6,7,8,9,10},
            xlabel={Iterations},
            ylabel={Sum-Rate [bps$/\SI{}{\hertz}$]},
            grid=both,
        ]
            \addplot[MUswmmsegmmpgdulfromy, TUMBeamerBlue, solid,mark=square,mark size=2pt,mark indices = {1,50,100,150,200,250,300}]
                table[x=iter,y=SWMMSE_GMM_observation, col sep=comma] {gmm_measurement/K8_B6/sumrate_10dB.csv};
                \addlegendentry{GMM samples, $\mby$};
                

            \addplot[MUswmmsegmmpgdulfromy, TUMBeamerGreen, solid,mark=square,mark size=2pt,mark indices = {1,50,100,150,200,250,300}]
                table[x=iter,y=SWMMSE_GMM_observation, col sep=comma] {gmm_measurement/K8_B6/sumrate_10dB_toep.csv};
                \addlegendentry{tGMM samples, $\mby$};
                

            \addplot[MUswmmsegmmpgdulfromy, TUMBeamerBlue, solid,mark=o,mark size=2pt,mark indices = {1,50,100,150,200,250,300}]
                table[x=iter,y=RCI_GMM_observation, col sep=comma] {gmm_measurement/K8_B6/sumrate_10dB_rciReg.csv};
                \addlegendentry{GMM, $\mby$};
                

            \addplot[MUswmmsegmmpgdulfromy, TUMBeamerGreen, solid,mark=o,mark size=2pt,mark indices = {1,50,100,150,200,250,300}]
                table[x=iter,y=RCI_GMM_observation, col sep=comma] {gmm_measurement/K8_B6/sumrate_10dB_toep_rciReg.csv};
                \addlegendentry{tGMM, $\mby$};
                

            \addplot[MUlloydpgdulgmmulall,dash dot,mark=x,mark size=2pt,mark indices = {1,50,100,150,200,250,300}]
                table[x=iter,y=RCI_DFT_hGMM,col sep=comma] {gmm_measurement/K8_B6/sumrate_10dB_rciReg.csv};
                \addlegendentry{\pltMUdftgmmulall};
            \addplot[MUlloydpgdulgmmulall,dash dot,mark=Mercedes star,mark size=2pt,mark indices = {1,50,100,150,200,250,300}]
                table[x=iter,y=RCI_DFT_hGMM,col sep=comma] {gmm_measurement/K8_B6/sumrate_10dB_toep_rciReg.csv};
                \addlegendentry{DFT, $\hat{\mbh}_{\text{tGMM}}$};
            \addplot[MUlloydpgdulomp,dash dot,mark=diamond,mark size=2pt,mark indices = {1,50,100,150,200,250,300}]
                table[x=iter,y=RCI_DFT_hOMP,col sep=comma] {gmm_measurement/K8_B6/sumrate_10dB_rciReg.csv};
                \addlegendentry{\pltMUdftomp};
            \addplot[MUlloydpgdulscov,dash dot,mark=triangle,mark size=2pt,mark indices = {1,50,100,150,200,250,300}]
                table[x=iter,y=RCI_DFT_hLMMSE,col sep=comma] {gmm_measurement/K8_B6/sumrate_10dB_rciReg.csv};
                \addlegendentry{\pltMUdftscov};

        \end{axis}
        
    \end{tikzpicture}
    \caption{The sum-rate over the number of iterations for a system with $B=6$ feedback bits, $J=8$ \acp{MT}, $n_p=8$ pilots, and $\text{SNR}=\SI{10}{dB}$.}
    \label{fig:meas_K8B6_overiter}
\end{figure}

In all the results so far, we have seen a superior performance of the generative modeling-based approaches ``\{GMM, tGMM\} samples, $\mby$''.
The significant performance gains were obtained through the \ac{SWMMSE} algorithm, fed with samples generated by the \ac{GMM}.
The \ac{SWMMSE} algorithm designs the precoders based on iterative updates.
Thus, in \Cref{fig:meas_K8B6_overiter}, we analyze the impact of the number of iterations on the sum-rate.
A few iterations are already enough to outperform all of the \ac{DFT}-based codebook approaches ``DFT, \{$\hhat_{\text{GMM}}$, $\hhat_{\text{tGMM}}$, $\hhat_{\text{OMP}}, \hhat_{\text{LMMSE}}$\}''. 
The generative modeling-based approaches ``\{GMM, tGMM\} samples, $\mby$'' require approximately $10$ or $20$ iterations to outperform their directional information-based counterparts ``\{GMM, tGMM\}, $\mby$,'' which utilize the \ac{RCI} method.
Thus, the \ac{GMM}-based feedback scheme allows for a trade-off between improved system performance and computational complexity and the associated latency for designing the precoders at the \ac{BS}.  

\section{Conclusion}

In this work, we utilized real-world measurement data to assess the performance of the recently proposed \ac{GMM}-based feedback scheme.
Despite exhibiting low complexity and enabling parallelization for inferring the feedback information, our experiments show that the \ac{GMM}-based feedback scheme, which is tailored to a communications environment, outperforms the well-established \ac{DFT}-based codebook solution in the considered system setups.
We conclude that the \ac{GMM}-based feedback scheme offers great potential for deployment in future wireless communication systems.

\balance
\bibliographystyle{IEEEtran}
\bibliography{IEEEabrv,biblio}
\end{document}